\documentclass[prb,twocolumn,showpacs]{revtex4} 
\usepackage{color}
\usepackage{graphicx}
\usepackage{amssymb,amsmath}
\begin{document}

\title{Melting of  stripe phases and its signature in the single-particle spectral function }

\author{Marcin Raczkowski}
\author{Fakher F. Assaad}
\affiliation{Institut f\"ur Theoretische Physik und Astrophysik,
             Universit\"at W\"urzburg, Am Hubland, D-97074 W\"urzburg, Germany}

\date{\today}

\begin{abstract}
Motivated by the recent experimental data [Phys. Rev. B {\bf 79}, 100502 (2009)] 
indicating the existence of a pure stripe charge order over 
unprecedently wide temperature range in La$_{1.8-x}$Eu$_{0.2}$Sr$_x$CuO$_4$, 
we investigate the temperature-induced  melting of the metallic stripe phase.   
In spite of taking into account local dynamic correlations within a
real-space dynamical mean-field theory of the Hubbard model, we observe a
mean-field-like melting of the stripe order irrespective of the choice of  the 
next-nearest neighbor hopping.  The temperature dependence of the single-particle spectral function
shows  the stripe induced formation of a flat band around the antinodal
points accompanied by the opening a gap in the nodal direction. 
\end{abstract}

\maketitle

In spite of intense theoretical studies over the two last decades, the debate
on the microscopic origin of the so-called stripe phases characterized by the
combined spin and charge order is far from closed.\cite{Vojta09}
So far, two main scenarios regarding the origin of
stripes have been put forward. In the weak-coupling scenario, stripe phases
emerge due to nesting properties of the Fermi surface and provide a
compromise between the superexchange interaction, which stabilizes the
long-range antiferromagnetic (AF) order in the parent Mott insulator, and the kinetic energy of
doped holes. In this scenario spin and charge orders occur at the same
temperature or charge stripe order sets in only after spin order has developed.
An alternative strong-coupling scenario comes from the Coulomb-frustrated phase
separation suggesting that  stripe formation is commonly charge
driven.\cite{Zach98} In this case the onset of charge order appears 
prior to spin order as the temperature is lowered. 
However, this scenario does not take into account spin fluctuations which
might prevent the spins from ordering at the charge-order temperature 
$T_{\rm CO}$.\cite{Duin98} 

Similarly, on the experimental side the origin of stripes remained up to now 
unclear. On the one hand, in the most widely studied stripe-ordered compounds 
La$_{1.6-x}$Nd$_{0.4}$Sr$_x$CuO$_4$ (LNSCO) \cite{Ichi00} and
La$_{2-x}$Ba$_x$CuO$_4$,\cite{Tran08} the charge order observed as
Bragg maxima at the wave vectors ${\pmb q}_c=(\pm 4\pi\epsilon,0)$ with
$\epsilon\simeq x$ for doping $x\lesssim1/8$  was found to set in at a \emph{higher}
temperature $T_{\rm CO}$ than the spin order signalized below $T_{\rm SO}$ by the onset of 
magnetic Bragg peaks at ${\pmb q}_s=\pi(1\pm 2\epsilon,1)$, thus
supporting the strong coupling scenario. On the other hand, the interval
between  $T_{\rm CO}$ and $T_{\rm SO}$ was found  not to exceed 15 K.  
In contrast, in the recent resonant soft x-ray studies of the stripe order in 
La$_{1.8-x}$Eu$_{0.2}$Sr$_x$CuO$_4$ (LESCO),\cite{Fink09} both temperatures were
found to be separated by 35 K. Those findings clearly exclude the nesting scenario of
the stripe formation in favor of the electronic origin.

In this paper we address melting of the metallic stripe order and look for a
signature of a pure charge-order above $T_{\rm SO}$. We consider the Hubbard model, 
\begin{equation}
H=-\sum_{ij,\sigma}t_{ij}c^{\dag}_{i\sigma}c^{}_{j\sigma} +
   U\sum_{i}n^{}_{i\uparrow}n^{}_{i\downarrow},
\label{Hubb}
\end{equation}
with a hopping amplitude $t$ ($t'$) between the (next-) nearest neighbor sites, respectively,
and with $U=10t$ standing for the on-site Coulomb interaction and solve it within
the real-space dynamical mean-field theory (RDMFT). It allows one to handle
the leading local part of dynamic correlations exactly.
Although the long-range Coulomb interaction might help to
  stabilize further the stripe order, the great success
of the previous RDMFT studies at $T=0$ was a proof that 
the correct treatment of the on-site interaction alone suffices to stabilize
experimentally observed metallic stripes in LNSCO.\cite{Fleck} 
Following Ref.~\onlinecite{Fleck}, we  decompose the square lattice into 
$L_u$ stripe supercells with $\mu=1\ldots L_c$ orbitals so that the self-energy
${\pmb \Sigma}_{\sigma}( {\pmb K},i \omega_m)$ and noninteracting Green's
function ${\pmb G}_{0} ({\pmb K},i \omega_m)$ become $L_c \times L_c $
matrices, where the wave vectors ${\pmb K}$ span the folded Brillouin zone (BZ). 
Furthermore, similarly to the standard DMFT,\cite{DMFT}
the RDMFT approximation neglects the  ${\pmb K}$ dependency of the
self-energy while allowing for its spatial dependency, i.e.,
${\pmb \Sigma }_{\sigma}({\pmb K},i \omega_m) \equiv {\Sigma}_{\mu\nu,\sigma}(i \omega_m)\delta_{\mu\nu} $.  
Next, the lattice model (\ref{Hubb}) is mapped onto a set of effective quantum
impurities subject to a dynamic bath ${\pmb {\cal G} }_{0,\sigma}(i \omega_m)$
that has to be determined self-consistently. Here we use a Hirsch-Fye
quantum Monte Carlo (QMC) impurity solver; it not only
provides access to finite temperatures but also has the advantage of simulating an effectively
infinite bath.

In particular, for a given site-diagonal element of the bath Green's function 
${\cal G}_{0,i\sigma}(i \omega_m)$, chosen to impose an initial stripe
configuration, we use $L_c$ times the QMC solver to obtain the interacting
Green's function ${\cal G}_{i\sigma} (i\omega_m)$ and then  we extract
through the local Dyson equation the corresponding self-energy,  
$\Sigma_{i\sigma} (i \omega_m)= {\cal G}^{-1}_{0,i\sigma}(i \omega_m) - {\cal G}^{-1}_{i\sigma} (i
\omega_m)$. Next, we compute a new bath Green's function using the
self-consistency condition in a matrix form:\cite{Note} 
${ \pmb {\cal G} }^{-1}_{0,\sigma}(i \omega_m) = {\pmb {\cal G}}^{-1}_{\sigma}(i\omega_m)\delta_{\mu\nu} 
+{\pmb \Sigma}_{\sigma} (i\omega_m)$, 
with ${\pmb {\cal G}}_{\sigma}(i\omega_m) =
\tfrac{1}{L_u} \sum_{\pmb K} \tfrac{1} { {\pmb G}_0^{-1}({\pmb K},i \omega_m) - {\pmb \Sigma}_{\sigma}
         (i \omega_m) }$.
It couples all the impurity sites and hence the actual charge density at a given
site depends effectively on the charge distribution at other sites in the stripe phase. 
The procedure is repeated till convergence is reached.  Finally, we represent 
the lattice Green's function, 
${\pmb G}^{-1}_{\sigma}( {\pmb K} , i \omega_m)=
           {\pmb G}_0^{-1}( {\pmb K} , i \omega_m) - 
       {\pmb \Sigma}_{\sigma} ( i \omega_m)  $, 
in the original BZ as follows,     
$g_{\sigma}({\pmb k}, i \omega_m ) = 
\tfrac{1}{L_c} \sum_{\mu,\nu = 1}^{L_c} e^{i {\pmb k}\left( {\pmb a}_\mu -
  {\pmb a}_\nu\right) } {\pmb G}_{\sigma}( {\pmb K} , i \omega_m) $,
and by rotation to the real-frequency axis via stochastic analytical continuation
method,\cite{Beach04}  extract the ${\pmb k}$-resolved single-particle spectral function 
$A({\pmb k},\omega)=-\tfrac{1}{\pi}{\rm Im} \sum_{\sigma}g_{\sigma}({\pmb k}, \omega )$.

\begin{figure}[t!]
\begin{center}
\includegraphics[width=0.43\textwidth]{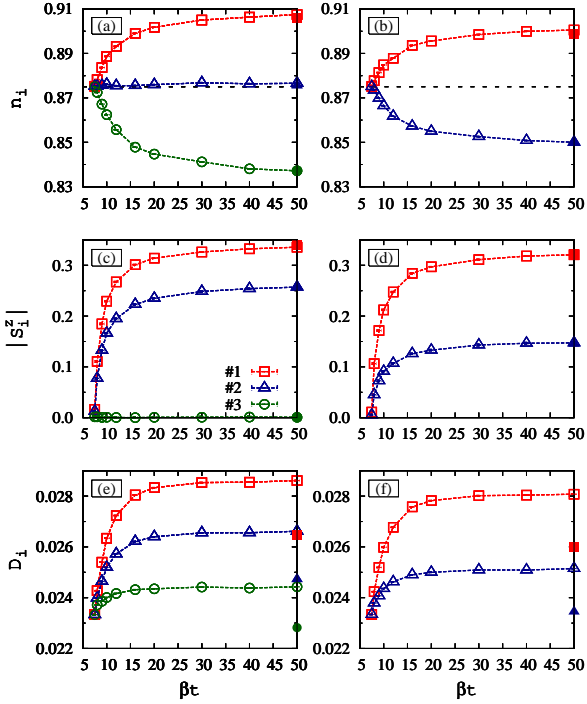} 
\end{center}
\caption
{(Color online)
  Temperature dependence of the: (a,b) local charge density $n^{}_{\rm i}$,
  (c,d) magnetization $|S^{z}_{\rm i}|$, as well as (e,f) double occupancies
  $D^{}_{\rm i}$ at the inequivalent sites in the SC (left) and BC (right) 
  stripe phase at doping $x=1/8$. Filled symbols at $\beta t=50$ were obtained
  with $t'=-t/3$. 
}
\label{obser}
\end{figure}

We begin by showing in Fig.~\ref{obser} the temperature evolution of the local charge 
$n^{}_{\rm i}=\sum_{\sigma}\langle c^{\dag}_{i\sigma}c^{}_{i\sigma}\rangle$  and magnetization 
$|S^{z}_{\rm i}|=\tfrac{1}{2}|\langle n_{i\uparrow} - n_{i\downarrow}\rangle|$ densities  
as well as changes in double occupancies $D^{}_{\rm i}=\langle n_{i\uparrow}
n_{i\downarrow}\rangle$ in the two possible vertical stripe patterns at doping
$x=1/8$: (i) site-centered
(SC) one with enhanced hole density at the nonmagnetic domain walls (DW) that
separate AF spin domains and (ii) bond-centered (BC) one consisting of
hole-rich ladders with a weak ferromagnetic order on the rungs separating AF
ladders. In fact, using the symmetry of
the self-energy in the spin sector, it is possible to reduce the number of
the impurity sites that must be solved down to  3 (2) sites in the SC (BC)
stripe phase, respectively, in a magnetic stripe unit cell with $L_c=8$ atoms.     
Guided by the experimental data one can expect  a complex
process of melting of the stripe order,  possibly via an intervening phase with
a pure charge order.\cite{Fink09}  We find instead that the spin and charge orders disappear
simultaneously at $\beta t\simeq 7$ and the overall melting is reminiscent of disappearance of the 
mean-field stripe phase in the overdoped regime.\cite{Racz05} This follows
from Fig.~\ref{obser}  where one observes a simultaneous increase in the  charge $n^{}_{\rm i}$ 
and spin  $S^{z}_{\rm i}$ order parameters on decreasing $T$.\cite{Riera,Lan}
A possible source of discrepancy  between the numerical and experimental data
lies in the absence of spatial 
fluctuations inherent to the RDMFT approach. Indeed,  the physics  of the 
proximity to  phase separation in the cuprates is not correctly captured since
the later instability corresponds to a long-wavelength instability. 
Alternatively, the melting of the stripe order in the layered cuprates might well be
controlled by the interplane coupling allowing for a phase transition to
occur at a finite-temperature. Indeed, it is not evident that the 
inter-plane coupling affects the charge and spin orders in the same way.
We also note that the enhancement of the 
double occupancies in the low $T$-regime gives  insight into the mechanism
of the stripe formation. The  kinetic energy gain acts as a 
driving force behind the stability of charge stripes along which doped
holes can propagate coherently as seen in the corresponding spectral functions (see later). 

\begin{figure}[t!]
\begin{center}
\includegraphics[width=0.43\textwidth]{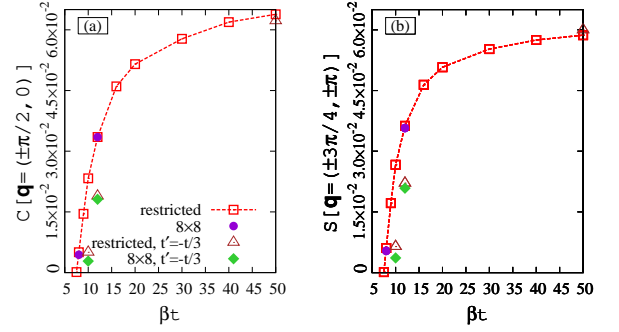} 
\end{center}
\caption
{(Color online)
  Intensity of the: (a) charge $C({\pmb q})$ and (b) spin $S({\pmb q})$ structure factor
  maxima at the wave vectors corresponding to the SC stripe
  phase with the charge (spin) unit cell containing 4 (8) atoms, respectively.  
  For clarity, $C({\pmb q})$ was multiplied by a factor of 4. 
}
\label{corel}
\end{figure}

To test the robustness of the stripe phase and exclude possible
stripe meandering into the AF domains we release all the constraints and 
solve self-consistently an $8\times 8$ stripe supercell at $\beta t=8$ and 12
where the use of the QMC solver albeit time expensive is nevertheless feasible. Further,
in order to facilitate the analysis, it is convenient to compare the intensity
of the cusps in the charge    
$C({\pmb q})= \tfrac{1}{N}\sum_{\pmb r}e^{{\pmb q}\cdot{\pmb r}}\langle n_0\rangle\langle n_{{\pmb r}}\rangle$
and spin 
   $S({\pmb q})= \tfrac{1}{N}\sum_{\pmb r}e^{{\pmb q}\cdot{\pmb r}}\langle
S^z_0\rangle\langle S^z_{{\pmb r}}\rangle$
structure factors at the appropriate wave vectors ${\pmb q}_c=(\pm\pi/2,0)$ and
${\pmb q}_s=(\pm3\pi/4,\pm\pi)$.
Again, as shown in Fig.~\ref{corel} we find that the  temperature dependence of
both  structure factors is very similar. 
Therefore, we conclude that relaxing the constraint has no overall effect even in the vicinity of the phase transition 
where thermal fluctuations are important.

\begin{figure}[t!]
\begin{center}
\includegraphics[width=0.3\textwidth]{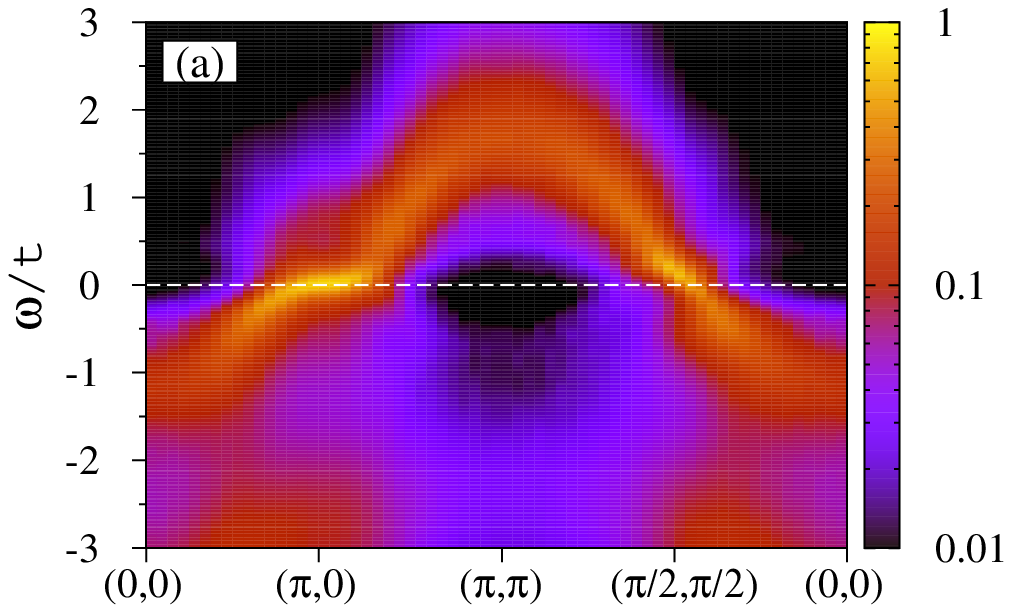} \\
\includegraphics[width=0.3\textwidth]{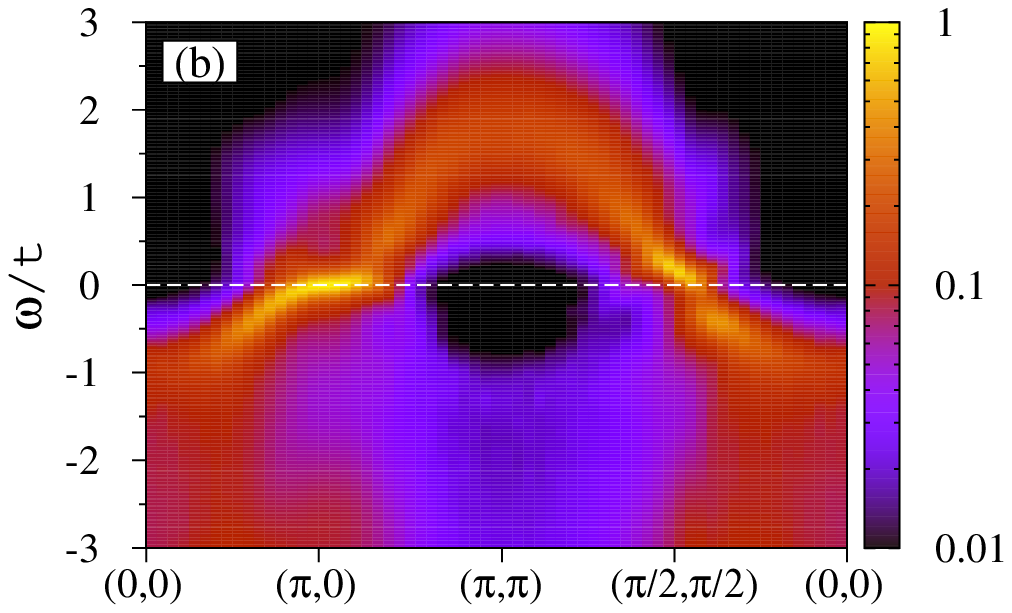}  \\
\includegraphics[width=0.3\textwidth]{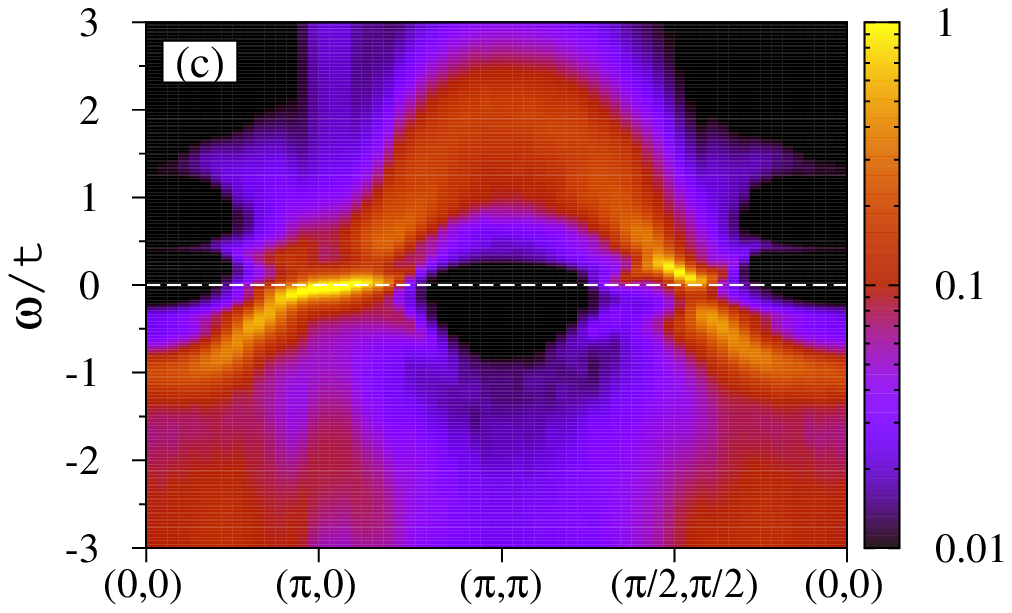} \\ 
\includegraphics[width=0.3\textwidth]{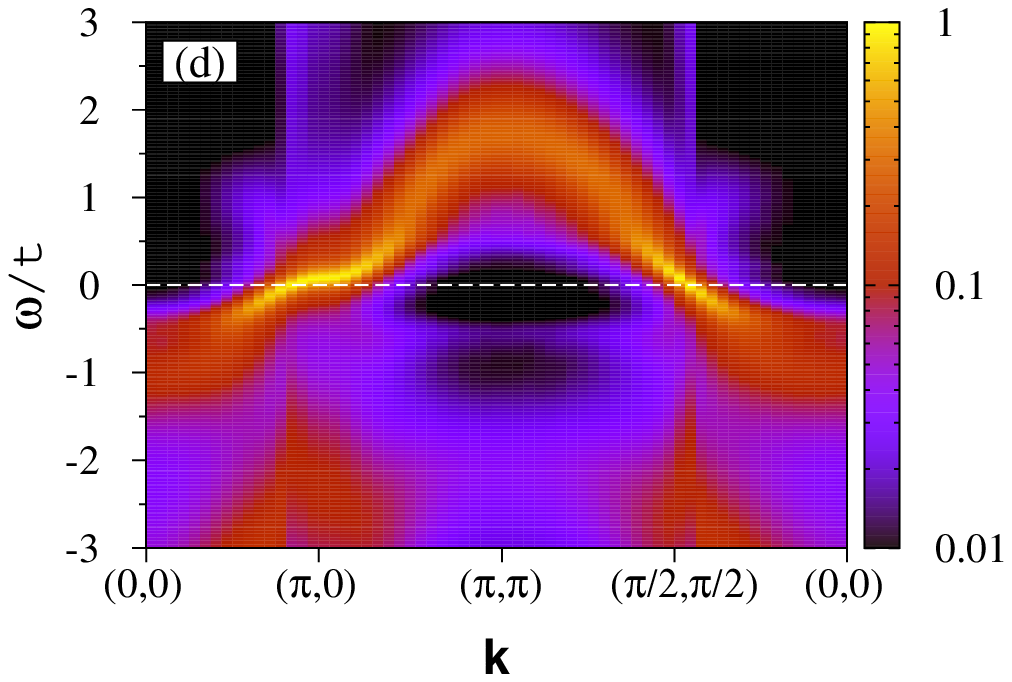}
\end{center}
\caption
{(Color online) Evolution of the low-energy part of the single-particle spectral function $A({\pmb k},\omega)$
  in the SC stripe phase: (a) $\beta t= 12$, (b) $\beta t= 16$, and (c)
  $\beta t= 50$; the SC and BC stripe phases are dynamically indistinguishable. 
  Panel (d) shows $A({\pmb k},\omega)$ in the PM phase at $\beta t= 50$.
}
\label{Ak}
\end{figure}

\begin{figure}[t!]
\begin{center}
\includegraphics[width=0.43\textwidth]{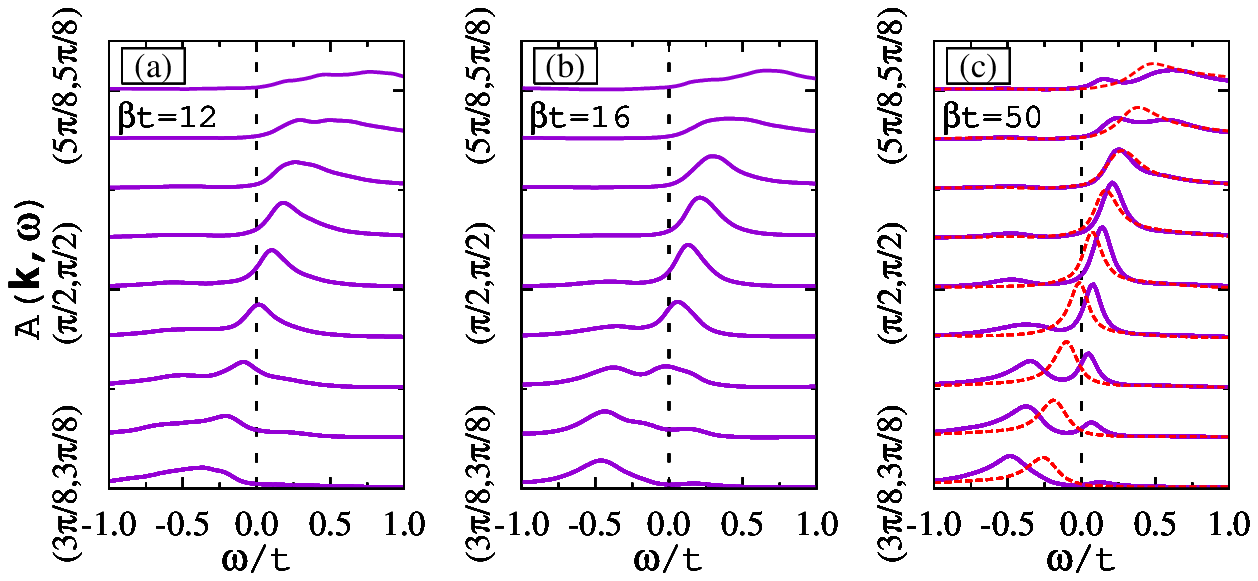}\\ 
\includegraphics[width=0.43\textwidth]{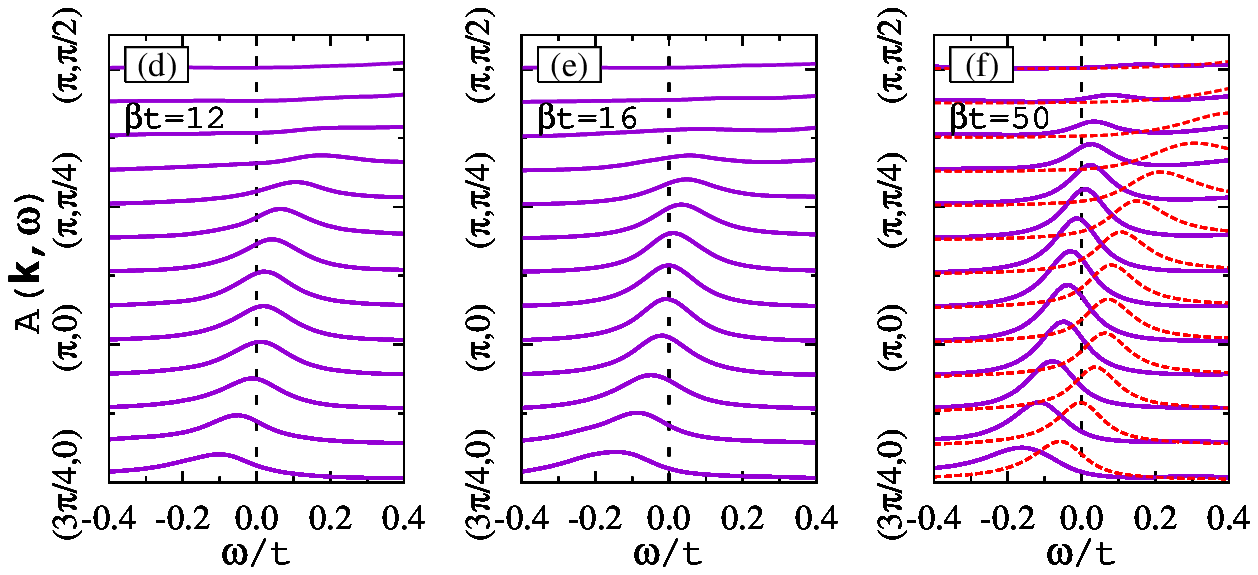}
\end{center}
\caption
{(Color online) Temperature dependence of the low-energy part of the spectral
  function $A({\pmb k},\omega)$
  in the SC stripe phase in the vicinity of: (a-c) the nodal $S=(\pi/2,\pi/2)$ 
  and (d-f) antinodal $X=(\pi,0)$  points. For comparison, dashed line in
  panels (c) and (f) shows $A({\pmb k},\omega)$ in the PM phase. 
}
\label{Ak_S}
\end{figure}

\begin{figure}[t!]
\begin{center}
\includegraphics[width=0.43\textwidth]{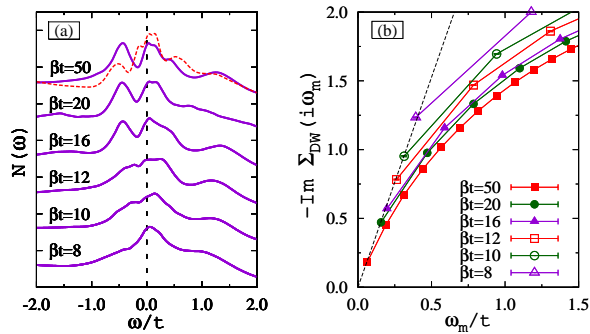}
\end{center}
\caption
{(Color online) (a) Temperature dependence of the density of states $N(\omega)$ in the SC stripe phase
  and (b) imaginary part of the self-energy at the DW.  Dashed line in
  panel (a) shows $N(\omega)$ in the PM phase while in panel (b) depicts the result of a linear fitting of data points at
the smallest Matsubara frequency $\omega_m=\pi T$.
}
\label{dos}
\end{figure}

The signature of the stripe  order emerges predominantly in the low-energy features of the 
single-particle spectral function $A({\pmb k},\omega)$. Figs.~\ref{Ak}(a-c) and
\ref{Ak_S} plot  this quantity as a function of temperature. 
Let us first concentrate on the nodal direction of the BZ.  Here, we note that thermally disordered stripes 
below $\beta t=12 $ (albeit clearly seen in the order
parameters in Fig.~\ref{obser}) leave $A({\pmb k},\omega)$ in the nodal
direction almost intact. These findings are consistent with
Ref.~\onlinecite{nodal} suggesting that the nodal $A({\pmb k},\omega)$ in the stripe system may arise in the
presence of disorder or due to fluctuations of stripes facilitating
penetration of the AF domains by holes. In contrast, a further reduction of temperature strengthens stripe
order and modifies its spectral properties. Indeed, already at
$\beta t=16$  one finds along the nodal direction two separated features originating from
the single quasiparticle (QP) peak. It results in the depletion of $A({\pmb k},\omega)$ at the Fermi
level $\varepsilon_F$.  Consequently, as depicted in Fig.~\ref{dos}(a), the QP peak seen at
$\varepsilon_F$ at $\beta t=8$ in the density of states $N(\omega)$ is superseded by a
pseudogap that opens up at $\beta t=16$ just below $\varepsilon_F$.  
Moreover, when moving towards  the $S=(\pi/2,\pi/2)$ point in Fig.~\ref{Ak_S}, spectral weight is transferred gradually
from the lower to the upper QP peak whose maximum  is far above
$\varepsilon_F$ at the $S$ point. Hence, in agreement with the previous $T=0$
RDMFT studies of the stripe order predicting opening a gap for charge
excitations along the diagonal direction,\cite{Fleck} it
seems that at the RDMFT level, well developed metallic stripes are incompatible
with the observed in La$_{2-x}$Sr$_x$CuO$_4$ (LSCO) and LNSCO nodal
quasiparticles.\cite{Zhou01} Altogether, noting that the measured nodal
$A({\pmb k},\omega)$ is more intense in LSCO (possible \emph{dynamic} stripes) than in LNSCO
(\emph{static} stripes) and comparing with the temperature dependence of  $A({\pmb k},\omega)$
shown in Fig.~\ref{Ak_S}, we confirm that the intensity of the nodal $A({\pmb
  k},\omega)$ provides a measure of the stripe \emph{disorder}.

Similarly to   the case of the  nodal direction,  it is only above $\beta t=12$ 
that we  observe the formation of a flat band   near the antinodal $X=(\pi,0)$
 point located just below $\varepsilon_F$. This flat  band can originate from 
the renormalization of the effective mass due to the 
frequency dependence of the self-energy -- as observed in the DMFT studies of  models
of heavy-fermions\cite{Racz10} -- or from its spatial
dependence.  To address this issue, we calculate $A({\pmb k},\omega)$ in
the uniform paramagnetic (PM) phase at the same doping $x=1/8$. As depicted in Figs.~\ref{Ak}(d) and \ref{Ak_S}, neither
the flat QP band at the $X$ point develops  nor the gap around the $S$ point opens
up in this case. Moreover, the Matsubara axis mass-renormalization factor
i.e., derivative of the imaginary part of the self-energy taken at the DW
w.r.t. the smallest Matsubara frequency
$\tfrac{-\partial Im\Sigma_{\rm DW}(i\omega_m)}{\partial\omega_m}|_{\omega_m=\pi T}$  does not
change across the stripe energy scale $\beta t\simeq 16$  [see Fig.~\ref{dos}(b)]
at which visible changes in $A({\pmb k},\omega)$ develop.
Hence, it is the site-dependence of the self-energy which is
responsible for  the substantial differences in the angle-resolved and integrated spectral functions between 
the PM and  stripe phases in the low-$T$ limit.

Finally, let us briefly address the influence of the next-nearest-neighbor
hopping $t'$ on the melting of the stripe order. On the one
hand, Fig.~\ref{obser} shows that the charge and spin profiles at our lowest
temperature $\beta t=50$ are only weekly affected by a finite $t'=-t/3$. On
the other hand, a destabilizing function  of $t'$ is seen as a strong reduction 
of $D_i$ which signalizes reduced kinetic energy gain. The stripe order
becomes more fragile and vanishes already at $\beta t\simeq 9$. While we have again not found a
spontaneous tendency towards stripe meandering in the $8\times 8$ unit cell,  one
nevertheless observes enhanced discrepancies w.r.t. the $t'=0$ case
indicating  the frustrating role of $t'$.

In summary, our  RDMFT calculations reveal a strong temperature dependence of $A({\pmb k},\omega)$ in the stripe phase. 
The dominant contribution to the electron mass renormalization  at the antinodal points  originates  from
the spatial dependence of the self-energy. Furthermore, it is only well below the transition temperature 
that a charge gap opens in the  nodal direction. 
The melting of the  stripe phase is mean-field-like since both charge and spin modulations vanish 
simultaneously. Within the RDMFT this was shown to be a robust result 
with respect to both $t'$ and the size of  the stripe supercell.  The discrepancies with the experimental situation  in
LESCO, in which a two-step melting is observed,  can be understood  in terms of the absence of spatial 
fluctuations inherent to the RDMFT approach. Additional argument comes from the recent dynamic cluster
approximation studies  showing that the short-range AF correlations in
the PM phase are sufficient for the occurrence of phase separation.\cite{dca} Hence, it
would be interesting to investigate whether the tendency towards phase 
separation can develop into a pure charge stripe order seen in LESCO.     
While other terms such as the long-range Coulomb interaction and interplane
coupling might also be necessary, a systematic improvement of the RDMFT aimed
at capturing spatial fluctuations is presently under progress.

\begin{acknowledgments}
We thank the LRZ-M\"unich and the J\"ulich Supercomputing center for a generous 
allocation of CPU time. M.R. acknowledges support from the Alexander von Humboldt Foundation. 
F.F.A. thanks the DFG for financial support.
\end{acknowledgments}


\begin{thebibliography}{10}
\bibitem{Note} Note that the Kronecker
  delta $\delta_{\mu\nu}$ is necessary to relate the diagonal elements of the 
  Green's function ${\pmb {\cal G}}_{\sigma}(i\omega_m)$ to the local Green's
  function ${\cal G}_{i\sigma} (i\omega_m)$.
\end{thebibliography}

\end{document}